# Reaction method control of impurity scattering in C-doped $MgB_2$, proving the role of defects besides C substitution level

S K Chen[1], K Y Tan[1], A S Halim[1], X Xu[2], S De Silva[2], W K Yeoh[2,3], S X Dou[2], A Kursumovic[4] and J L MacManus-Driscoll[4]

[1]Department of Physics, Faculty of Science, Universiti Putra Malaysia, 43400 Serdang, Selangor, Malaysia.

[2]Institute for Superconducting & Electronic Materials, University of Wollongong, North Wollongong, New South Wales 2500, Australia.

[3]Australian Centre for Microscopy & Microanalysis, University of Sydney, Sydney, New South Wales 2006, Australia.

[4]Department of Materials Science and Metallurgy, University of Cambridge, Pembroke Street, Cambridge CB2 3QZ, United Kingdom.

E-mail: chensk@upm.edu.my

**Abstract.** In this study, Si and C were incorporated into polycrystalline $MgB_2$ via *in situ* reaction of Mg and B with either SiC or with separate Si and C (Si+C). The electrical transport and magnetic properties of the two series of samples were compared. The corrected resistivity at 40K, $\rho_A(40K)$, is higher for the SiC reacted samples regardless of carbon (C) substitution level, indicating larger intragrain scattering because of the simultaneous reaction between Mg and SiC and carbon substitution during the formation of $MgB_2$. In addition, because of the cleaner reaction route for the SiC reacted samples, the calculated active area that carries current, $A_F$, is twice that of the (Si+C) samples. On the other hand, the upper critical field, $H_{c2}$, was similar for both sets of samples despite their different C substitution levels which proves the importance of defect scattering in addition to C substitution level. Hence, the form of the precursor reactants is critical for tuning the form of $H_{c2}(T)$.

# 1. Introduction

SiC appears to be one of the promising dopants which has been used widely to enhance the critical current density, $J_c$ as well as upper critical field, $H_{c2}$, of the MgB$_2$ superconductor [1-5] thus enabling the generation of high magnetic fields. This is of great interest for mid field (2 – 5T) applications and Magnetic Resonance Imaging (MRI) magnet operating at 20K. In fact, doping with silicon (Si) or carbon (C) alone has also been found to improve the electromagnetic properties of MgB$_2$ greatly as a result of impurity scattering [6].

For doping with Si, the field dependent $J_c$ and $H_{c2}$ are enhanced and the superconducting transition temperature, $T_c$ is only decreased by about 2K with as much as 10 wt.% dopant additions compared with the pure sample [7, 8]. Detailed Rietveld analysis of the XRD data shows no sign of Si substitution into the lattice of MgB$_2$. Hence, it is believed that the nano sized precipitates consisting of Si-related bi-products are distributed within the MgB$_2$ matrix serving as effective pinning centres. Doping MgB$_2$ with C reduces the *a*-axis systematically suggesting substitution of C on the B site [9] as the C-B bond is shorter than that of B-B [10, 11]. Although C doping is thought to predominantly disturb the σ-band, the increase in scattering for π-band is even more rapid [12]. Compared with Si, C enhances $H_{c2}$ while it reduces the anisotropy of this critical parameter with respect to the different crystallographic directions of MgB$_2$ [13]. A large $H_{c2}(0K)$ exceeding 32T and 60T has been reported for C doped bulk [14, 15] and thin films [16], respectively. The value of $H_{c2}$ for the latter is much larger than that of NbTi and Nb$_3$Sn bulks [16].

It has been demonstrated that MgB$_2$ doped with SiC has the weakest field dependence of $J_c$ at 20K compared with Si or C [1, 4, 17]. Such a superior field performance is an indication of combined effect of C doping and flux pinning by the defects and nano precipitates [4]. The C source for doping is typically from the reaction between SiC and Mg forming Mg$_2$Si [18], leaving behind C which in presence of B, reacts with Mg to form Mg(B$_{1-x}$C$_x$)$_2$. The remaining Si does not substitute into the lattice of MgB$_2$ [7, 19]. As a result of C substitution, the anisotropy in $J_c$ is reduced [20].

The availability of free Si and C in the SiC doped polycrystalline $MgB_2$ [21] suggests that doping with SiC and separate (Si+C) may have a similar effect. The aim of this work is to verify how these different reaction routes influencing the doping and defect levels and hence the superconducting properties of $MgB_2$. Two series of samples reacted with SiC and separate Si and C (Si+C), respectively, were prepared and their electrical transport and magnetic properties compared. To optimise the reaction processes, both sample sets were sintered at 650ºC (around the melting point of magnesium) and 850ºC, and the resulting grain connectivity [22] and C substitution levels [23] were compared.

**2. Experimental Details**

Polycrystalline samples were made by the direct *in situ* reaction method. Starting raw materials are magnesium (Tangshan, 99%), amorphous boron (Pfaltz & Bauer, 99%), silicon carbide (Nano-Amor, 15nm, 99+%), silicon (Nano-Amor, 50 nm, 99+%) and carbon (Nano-Amor, 30 nm, 99+%) powders. Appropriate amounts of magnesium (Mg) and boron (B) (molar ratio of 1:2) were mixed with 3.0 weight percentage (wt.%) of silicon carbide (SiC) and hand-ground using a pestle and mortar for about 2 h. The mixed powders were then cold pressed into pellets of approximately 13 mm diameter and 2 mm thickness using a hydraulic press at pressure of 5 tons. The pellets were sealed into a stainless steel tube and later loaded into a tube furnace for sintering. For comparison, another set of samples were prepared in a similar way using 3.0 wt.% of individual silicon and carbon (Si+C) nano powders. Samples reacted with 3.0 wt.% SiC and Si+C were chosen for the present study because they have a high critical current density [24]. Undoped $MgB_2$ pellets were also prepared as control samples. Sintering was undertaken at 650ºC and 850ºC for 1 h with heating and cooling rates of 10ºC/min. The tube of the furnace was clamped at both ends and subsequently flushed with argon gas for about 30 min prior to sintering. Argon gas flow was maintained during the entire heat treatment. Except for the sintering process, the rest of the experimental procedure was carried out in air. The details of sample preparation and their identity are summarised in table 1. Phase formation of the samples was checked using the X'Pert Pro Panalytical

PW3040 MPD X-ray Diffractometer with Cu-k$_\alpha$ radiation source. The $\theta$ - $2\theta$ scanning mode was used over a range of angles from 20º to 80º with step size of 0.02º. The microstructures of the fractured surface of the pellets were observed using a Jeol 6340F field emission gun scanning electron microscope (FEG-SEM). Magnetic measurements on the bar-shape samples (each with dimensions of approximately 1 mm × 1 mm × 2 mm) were carried out using a commercial Magnetic Property Measurement System (Quantum Design MPMS-XL). Magnetic moment versus temperature was measured after zero-field cooling and then by applying a field of 20Oe before warming the samples to normal state. For hysteresis loop measurements at 5K and 20K, the field was applied to the longest dimension of the samples. Critical current density, $J_c$ as a function of magnetic field, $H$, was calculated using the equation $J_c(H) = \Delta M(H) / [a(1 - a/3b)]$ where $2a$ and $2b$ ($a \leq b$) are sample dimensions for the cross section which is perpendicular to the applied field and $\Delta M$ is the width of the magnetization hysteresis loop [25]. Four-point electrical resistivity between 20 and 300K in applied magnetic field up to 13T was measured using a commercial Physical Properties Measurement System (Quantum Design PPMS).

## 3. Results and Discussion

*3.1 X-ray Diffraction*

Figure 1(a) and (b) show the x-ray diffraction (XRD) patterns of the samples sintered at 650ºC and 850ºC, respectively. Phase analysis was performed using the X'Pert HighScore Plus software. The majority of the peaks can be indexed as $MgB_2$ (ICSD reference no.: 98-000-9725) showing the dominance of this phase. Minority peaks associated with $MgB_4$, MgO and $Mg_2Si$ were also identified. The peak of $MgB_4$ phase (at around 35.7º) with a very low intensity could only be found in the samples sintered at 850ºC (figure 1(b)). The formation of $MgB_4$ is a result of more severe Mg evaporation giving rise to Mg deficiency [26]. However, the value of the relative intensity fraction [24] of $MgB_4$ is very small and not shown in table 1. The formation of MgO is expected in view of the powder handling in air during sample preparation. Moreover, the raw Mg powders may have already contained some MgO as Mg is reactive to oxygen even in air (Gibbs energy of formation for MgO at 298K is around -569 kJ/mol) [27].

For the pure sample sintered at 650ºC, the calculated relative intensity fraction for MgO is 4.6% compared with 8.5% for the sample sintered at 850ºC (table 1) indicating the latter was oxidised more severely. As shown in table 1, the calculated relative intensity fraction [24] of $Mg_2Si$ for the samples sintered at 650ºC is larger compared with that of the samples sintered at 850ºC. This is expected based on the fact that $Mg_2Si$ has a more negative Gibbs energy of formation at the lower temperature (-71.3 KJ·mol$^{-1}$ at 900K compared with -65.7 KJ·mol$^{-1}$ at 1100K) [27], thus making it form more easily. The presence of free Si in the (Si+C) samples accelerates the reaction between Si and Mg leading to a higher relative intensity fraction of $Mg_2Si$ as compared with the SiC samples

Rietveld refinement on the XRD data was performed using the same software (X'Pert HighScore Plus) in order to estimate the unit cell lattice parameters and strain. As shown in table 1, the *a*-axis decreases for both reaction precursors, indicating increasing levels of C substitution at B sites. The reduction in the *a*-axis is further enhanced thermodynamically by sintering the samples at 850ºC.

The C substitution level, $x$ for $Mg(B_{1-x}C_x)_2$ was estimated according to Ref. [28]. For samples sintered at 650ºC, $x$ is slightly higher for the samples reacted with SiC as expected [18] because more C is available as the reaction between SiC and Mg in forming $Mg_2Si$ is more favourable at lower sintering temperature [27]. Conversely, for samples sintered at 850ºC, $x$ is higher for the samples reacted with (Si+C) due to a higher solubility of free C at higher temperature [23]. At 650°C, the difference in $x$ between the SiC and (Si+C) samples is 0.0018, which is less (by around half) than the difference between the 850°C sintered samples (0.0035). Assuming a linear relationship between the *a*-axis and C substitution level as for $MgB_2$ single crystals [29], the estimated $x$ values for samples (Si+C)650C and (Si+C)850C are 0.008 and 0.015, respectively, compared with 0.0084 and 0.0211 (the lowest and highest $x$, respectively) as calculated in this work using x-ray diffraction data (table 1).

An expansion of the *c*-axis as a result of C substitution is noticeable and has also been reported elsewhere for C doped $MgB_2$ bulks [30] and thin films [31]. The estimated level of strain is larger for the samples sintered at 650ºC. This is related to the formation of a higher density of defects when $MgB_2$ forms at around the melting point of magnesium [32]. Higher sintering temperature leads to the relief of strain because of improved phase formation and crystallinity [22].

Based on the FEG-SEM images captured at several areas across the samples, we estimated the grain sizes to be 100 – 400 nm and 300 – 700 nm for 650°C and 850°C sintering, respectively. However, the grain size shows not clear dependence on the reaction precursor, whether it be SiC or (Si+C). The density calculated as the mass per unit volume of the sample is approximately 1.42 $g/cm^3$, which is about 55% of the theoretical density of $MgB_2$.

*3.2 Electrical Transport and Magnetic Properties*

Figure 2 shows the temperature dependence of resistivity measured at zero field. The plots were normalised to the resistivity at room temperature, $\rho_{300K}$. In general, the resistivity for 200 - 300K is fairly linear. The temperature dependence of resistivity in 40 – 200K for all the samples can be fitted by a power law $\alpha + \beta T^{\gamma}$. The value of $\gamma$ obtained by curve fitting is shown in table 2. The $\gamma$ for the SiC and (Si+C) samples is in the range 2.53 – 2.83. It has been reported that undoped polycrystalline $MgB_2$ samples show $\gamma \sim 3$ [33]. However, Chen et *al.* found that $\gamma$ varies within the range 2.17 – 2.43 depending on the nominal Mg content, $x = 1.0 – 1.5$ in $Mg_xB_2$ [34]. Dense polycrystalline $MgB_2$ has a lower $\gamma \sim 1.9 – 2.1$ [35, 36]. For single crystals, the values are $2.7 < \gamma < 2.8$ [37] and 3 [38]. Clearly, the values for our doped samples (2.53 – 2.83) are similar to single crystal values.

As shown in table 2, the residual resistivity ratio, *RRR* [39] is smaller for the samples reacted with SiC or (Si+C) as compared with that of the pure samples. The range of *RRR* is close to previously reported values for pure and SiC doped $MgB_2$ samples [1, 3]. In general, the smaller *RRR* means the dirtier the samples due to defects or disorder and impurities as has been found elsewhere in $MgB_2$ doped with SiC [1, 3], C [19, 29, 30], Al [40], samples irradiated with neutron [40] and post annealed in Mg vapour [1]. The *RRR* for "clean" undoped polycrystalline $MgB_2$ is ~ 15 [1] while for single crystals the *RRR* value is lower at ~ 5 [29, 37] or 7 [19, 29]. For the pure samples, lower sintering temperature (650ºC) will give a higher density of defects and this accounts for the slightly lower *RRR* compared with sintering at 850ºC. For the sample reacted with (Si+C) at 650ºC, its *RRR* is higher compared with that of the SiC sample possibly due to the lower C substitution level (table 1).

The corrected resistivity at 40K, $\rho_A(40K)$ [39] is also shown in table 2. Regardless of the sintering temperature, the larger $\rho_A(40K)$ for samples reacted with SiC indicates stronger intragrain scattering compared with the samples reacted with (Si+C). This can be correlated to the simultaneous reaction of Mg and SiC (forming $Mg_2Si$) followed by C substitution during the formation of $MgB_2$ which results in a high density of defects and disorder, in addition to impurities of mainly remnant $Mg_2Si$ within the grains [1, 18, 41]. Indeed, for the SiC reacted samples higher FWHMs of the (110) plane are observed compared with the samples reacted with (Si+C) prove the more distorted lattice structure for the former [32].

The calculated active area for supercurrent flow, $A_F$ [39] decreases from 0.085 for the pure sample to 0.056 and 0.027 for the samples reacted with SiC and (Si+C) at 650ºC, respectively (table 2). Upon increasing the sintering temperature to 850ºC, the $A_F$ increases to 0.107 for the pure sample but no pronounced change in the $A_F$ value is noticeable in the samples reacted with SiC and (Si+C). This shows that the improvement to the connectivity with temperature is hindered by the impurity phases which arise as a result of the reaction between the additives and Mg and B. The lower connectivity for the samples reacted with (Si+C) is linked to the presence of residual C and a higher fraction of $Mg_2Si$, which interrupt the grain connectivity [41, 42]. The larger active area for the SiC samples also agrees with the $\rho_A(40K)$ data supporting the fact that more defects and impurities are within the grains compared to the (Si+C) samples, rather than at grain boundaries.

The superconducting transition temperature, $T_c$ and the breadth of transition, $\Delta T_c$ estimated from figure 2 are shown in table 2. $T_c$ is defined as the peak of the first derivative of normalised zero field resistivity over temperature, $d(\rho/\rho_{300K})/dT$ while the breadth of transition, $\Delta T_c$ is defined as the difference between the $T_{c\text{-onset}}$ and $T_{c\text{-offset}}$ of the same plot ($\Delta T_c = T_{c\text{-onset}} - T_{c\text{-offset}}$) as shown in the inset of figure 2 [43]. As expected, the samples reacted with SiC and (Si+C) have a lower $T_c$ than the pure sample, due to the reduced hole density of states as a result of C substitution [44]. In general, $T_c$ decreases with C substitution level, $x$ (table 1 and 2).

Upon increasing the sintering temperature to 850°C, the $T_c$ for the SiC samples did not change much because the C substitution level, $x$, changed little, from 0.0102 to 0.0176. The difference in $T_c$ between the samples with the lowest ($x = 0.0084$) and highest ($x = 0.0211$) C substitution levels is 0.9K. This is slightly higher than the estimated value based on C doped $MgB_2$ filaments which is 0.7K [14]. As shown in table 2, a considerable broadening of $\Delta T_c$ is obvious for both the sample sets but this is more severe for the SiC set. Also, the broadening in $\Delta T_c$ is larger by ~ 0.7K for the samples sintered at 850ºC as a result of more vigorous reaction between the additives and Mg and B. This broadened $T_c$ coincides with the way $\rho_A(40K)$ changes (table 2) indicating appreciable influence of intragrain defects and impurities on the superconducting transition, this being much greater in the SiC samples than the (Si+C) samples.

The temperature dependence of resistivity was also measured at applied fields of 0.5 – 13T. A broadening of resistive transition in the presence of external magnetic fields is noticeable especially for the pure samples, as shown in figure 3 (only a few plots are shown here for ease of discussion). The in-field resistive broadening has been linked to thermally activated flux flow (TAFF) in bulk and textured $MgB_2$ films [45, 46]. In polycrystalline $MgB_2$ samples with randomly oriented grains, broadening of the resistive transition in field is amplified due to the anisotropic upper critical field [47]. As opposed to the $MgB_2$ bulks [33, 48] and clean films [49], the magnetoresistance effect for all the samples of this work is negligibly small. A similar effect has been observed in the dense $MgB_2$ bulks [35, 36, 50] supporting the fact that all of our samples are in the dirty limit, as expected [33, 48].

Figure 4 shows the upper critical field, $H_{c2}$ which was defined as the field at 90% of the resistive transition curve. The obtained $H_{c2}$ represents the maximum value for polycrystalline $MgB_2$ due to anisotropic nature of this critical parameter. The temperature dependent $H_{c2}$ curve for the SiC sample sintered at 850°C is the steepest amongst the samples. The $H_{c2}$ at 20.6K for this sample is 13T, a value which is comparable to that of 10 wt.% SiC doped $MgB_2$ samples [18].

Figure 5 summarises the difference in the physical parameters for the two different reaction routes and two different reaction temperatures. The plot is not intended to show the variation of parameters with $x$ since there are insufficient points, rather the clear differences between the properties and how they depend on both the reaction precursors and reaction temperatures. Firstly, we observe the different C substitution levels between the SiC and (Si+C) samples which is clearly manifest by the upward shift of the SiC data from the (Si+C) data in Fig. 5(a) - (c). It is also evident that $x$ increases with sintering temperature in both set of samples. Secondly, we observe that the SiC samples have a higher $\rho_A(40K)$, $\Delta T_c$ and $A_F$ (figure 5(a) – (c)) regardless of the sintering temperature, because the density of intragrain defects and impurities is higher and the distortion to the lattice structure is more severe while the density of intergrain defects and impurities is lower.

Finally, we compare the $H_{c2}$ values (figure 5(d)) of the samples. The $H_{c2}$ values were estimated at 20.6K since this is the lowest temperature at the highest field for the steepest $H_{c2}$ curve (sample SiC850C). A very interesting finding is that unlike $MgB_2$ reacted with C only [13-15], the $H_{c2}$ (at 20.6K) for our samples does not increase with increasing C substitution level. Hence, sample (Si+C)650C ($x = 0.0084$) has a higher $H_{c2}$ than sample SiC650C ($x =$ 0.0102). As discussed earlier, the higher density of defects, impurities and poorer crystallinity [32] rather than the precise C doping level is what dominates the scattering and hence $H_{c2}$ [51].

The slopes of the $H_{c2}$ versus $x$ graphs are opposite for the SiC and (Si+C) samples. The negative slope for the (Si+C samples) arises because the crystallinity and impurity level which dominate the scattering decrease. On the other hand, for the positive slope for the SiC sample giving the highest $H_{c2}$ for the SiC850C sample is because of the combined effect of it having a relatively higher C dopant level ($x = 0.0176$), as well as high defects concentration which act together to give strong scattering as indicated by the largest value of $\rho_A(40K)$ and $\Delta T_c$ (figure 5(a) and (b)). The combined defect landscape arises from the specific dual reaction mechanism as discussed already [18]. Overall, we are able to achieve similar $H_{c2}$ values as samples with higher $x$ values of 0.021 made by reaction with pure C [14].

Figure 6 shows the critical current density, $J_c$ versus field measured up to 7T at both 5 and 20K. There is a general improvement in $J_c(H)$ with carbon substitution level, $x$, consistent with literature [52]. However, the sample reacted with SiC at 850°C has the best $J_c(H)$ despite having a slightly lower C doping level than the next best sample (Si+C) reacted at 850°C. The result is explained by the sample having a larger active area for supercurrent in addition to both enhanced scattering and flux pinning from the higher density of intragrain defects. We note that we have also measured the hysteresis loops for the samples reacted with 10 wt.% of the respective dopants as well as the 3 wt.% studied in detail here, and we found no significant difference in $J_c$ (either at 5 or 20K). However, the 10 wt.% samples reacted with SiC at 850°C has a weaker field dependent $J_c$ at 20K compared to the same weight percentage of (Si+C) samples as reported previously [52].

## 4. Conclusions

The electrical transport and magnetic properties of the polycrystalline $MgB_2$ reacted with SiC and (Si+C) were compared. As a result of different reactivity between SiC and (Si+C) with Mg and B, the former leads to a higher intragrain scattering as well as a higher active area for current transport. While the SiC reacted samples have a lower C substitution levels than (Si+C) reacted samples, they have a higher density of defects which produces additional scattering, as well as having cleaner grain boundaries which impede current transport less across them. Hence, these samples have the highest $H_{c2}$ at around 20K as well as the best $J_c(H)$ performance at 5 and 20K. The optimum sample studied, both in terms of $H_{c2}$ and field dependence of $J_c$, was a SiC sample reacted at 850°C.

**Acknowledgements**

This work is supported by the Ministry of Higher Education Malaysia under the Exploratory Research Grant Scheme (ERGS). S. X Dou would like to thank Hyper Tech Research, Inc. and Australian Research Council for their support. JLM-D acknowledges the European Research Council (ERC) (Advanced Investigator grant ERC-2009-AdG-247276-NOVOX).

**Table 1.** Sample preparation conditions, relative intensity fraction of phases, lattice parameters and estimated C substitution level, $x$.

| Sample identity | Sample preparation condition | Relative intensity fraction of phase (%) | | $a$-axis (Å) | $c$-axis (Å) | $x$ in [Mg(B$_{1-x}$C$_x$)$_2$] |
| --- | --- | --- | --- | --- | --- | --- |
| | | MgO | Mg$_2$Si | | | |
| P650C | Mg + 2B sintered at 650ºC for 1 h | 4.67 | - | 3.0863(1) | 3.5248(2) | - |
| SiC650C | Reaction of Mg + 2B with 3.0 wt.% of SiC at 650ºC for 1 h | 5.20 | 6.49 | 3.0835(2) | 3.5258(3) | 0.0102 |
| (Si+C)650C | Reaction of Mg + 2B with 3.0 wt.% of (Si+C) at 650ºC for 1 h | 4.32 | 7.84 | 3.0831(4) | 3.5246(5) | 0.0084 |
| P850C | Mg + 2B sintered at 850ºC for 1 h | 8.50 | - | 3.0860(1) | 3.5247(2) | - |
| SiC850C | Reaction of Mg + 2B with 3.0 wt.% of SiC at 850ºC for 1 h | 5.29 | 1.79 | 3.0810(1) | 3.5262(2) | 0.0176 |
| (Si+C)850C | Reaction of Mg + 2B with 3.0 wt.% of (Si+C) at 850ºC for 1 h | 3.70 | 3.20 | 3.0806(2) | 3.5272(2) | 0.0211 |

**Table 2.** Resistivity properties, superconducting transition temperature and transition breadth.

| Samples | $\gamma$ | RRR | $\rho_A(40K)$ [$\mu\Omega\cdot cm$] | $A_F$ | $T_c$ [K] | $\Delta T_c$ [K] |
|---|---|---|---|---|---|---|
| P650C | 4.21 | 2.26 | 5.77 | 0.085 | 37.8 | 0.7 |
| SiC650C | 2.63 | 1.59 | 12.45 | 0.056 | 35.5 | 3.4 |
| (Si+C)650C | 2.53 | 1.72 | 10.21 | 0.027 | 36.0 | 2.1 |
| P850C | 2.63 | 2.43 | 5.10 | 0.107 | 37.9 | 0.7 |
| SiC850C | 2.83 | 1.54 | 13.56 | 0.058 | 35.7 | 4.1 |
| (Si+C)850C | 2.63 | 1.57 | 12.72 | 0.028 | 35.1 | 2.7 |

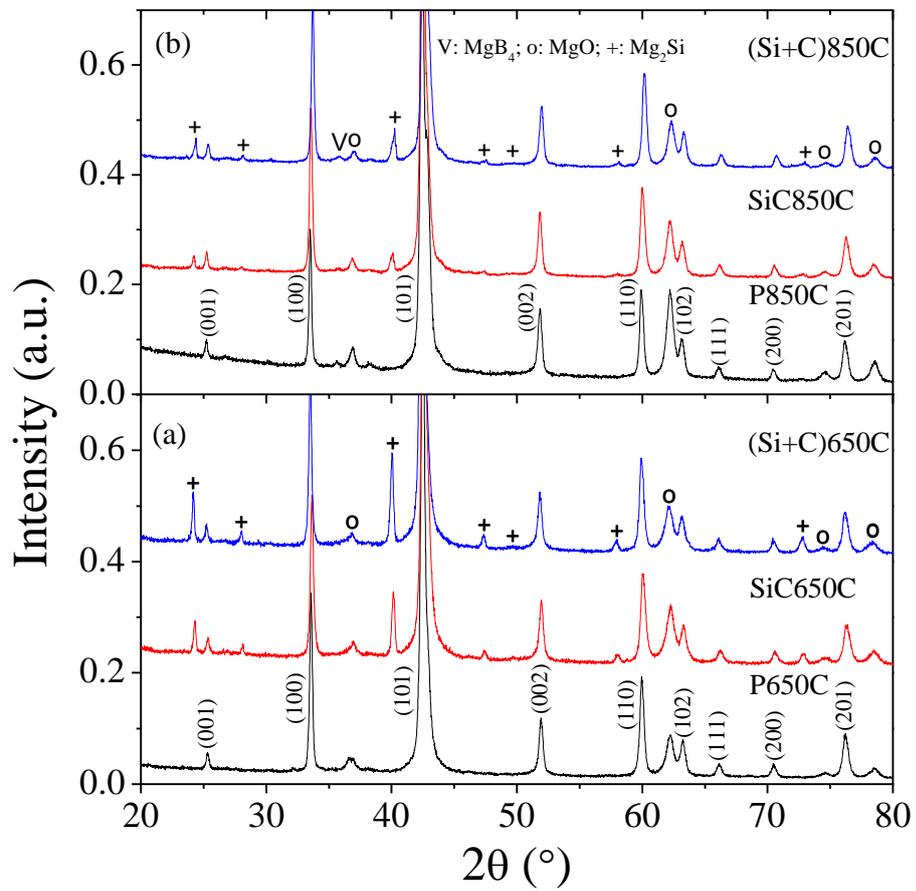

**Figure 1.** X-ray diffraction patterns for the pure, SiC and (Si+C) samples sintered at (a) 650°C and (b) 850°C.

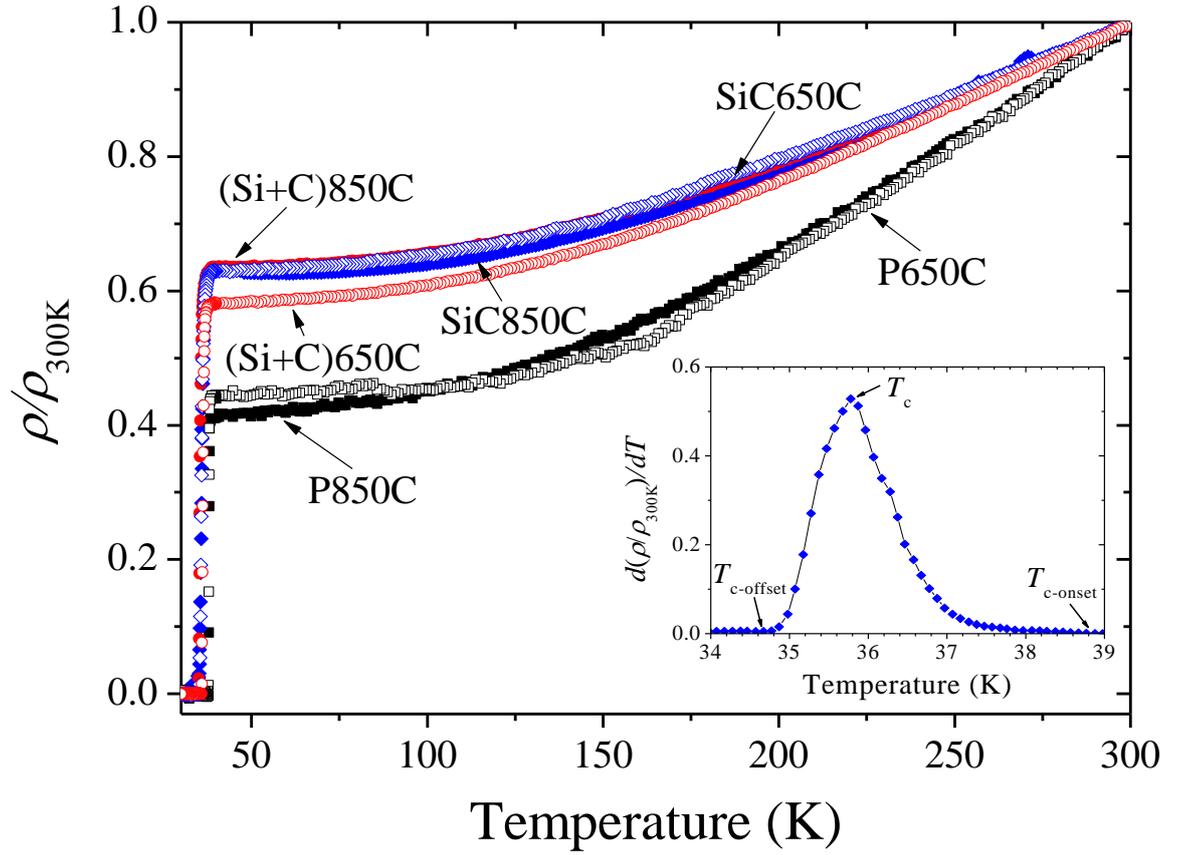

**Figure 2.** Normalised resistivity versus temperature plots in the range 20 – 300K. Inset: $T_c$, $T_{c\text{-onset}}$ and $T_{c\text{-offset}}$ as defined in the $d(\rho/\rho_{300K})/dT$ versus temperature plots.

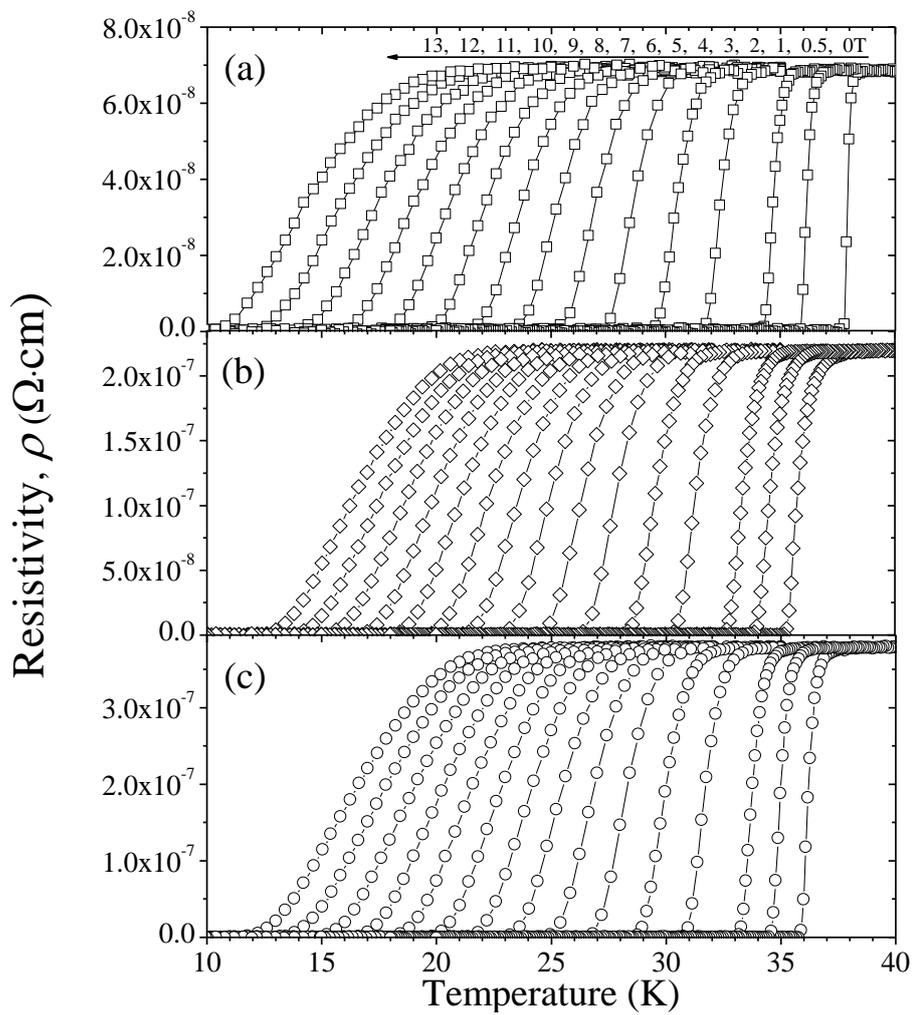

**Figure 3.** Resistivity versus temperature for the (a) pure (b) SiC and (c) (Si+C) samples sintered at 650°C. The applied magnetic fields are indicated on top of the plot in figure (a).

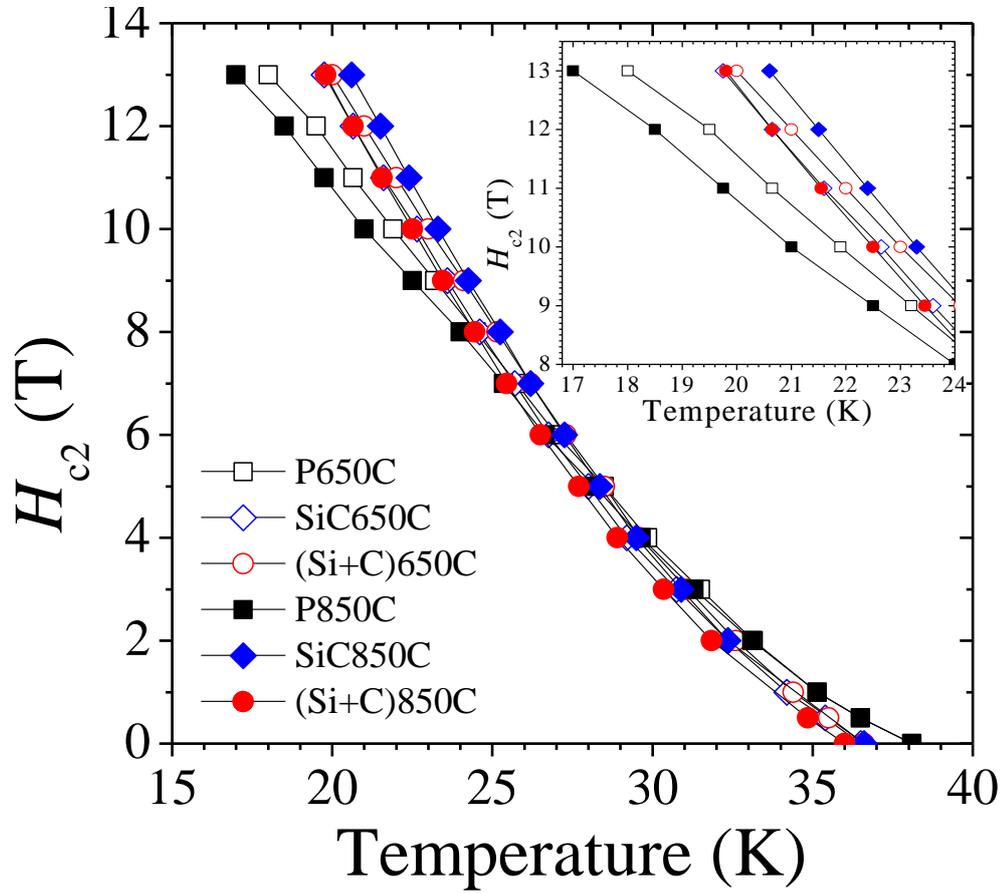

**Figure 4.** Temperature dependence of upper critical field, $H_{c2}$, measured up to 13T. Inset: $H_{c2}$ plots in a narrower temperature range.

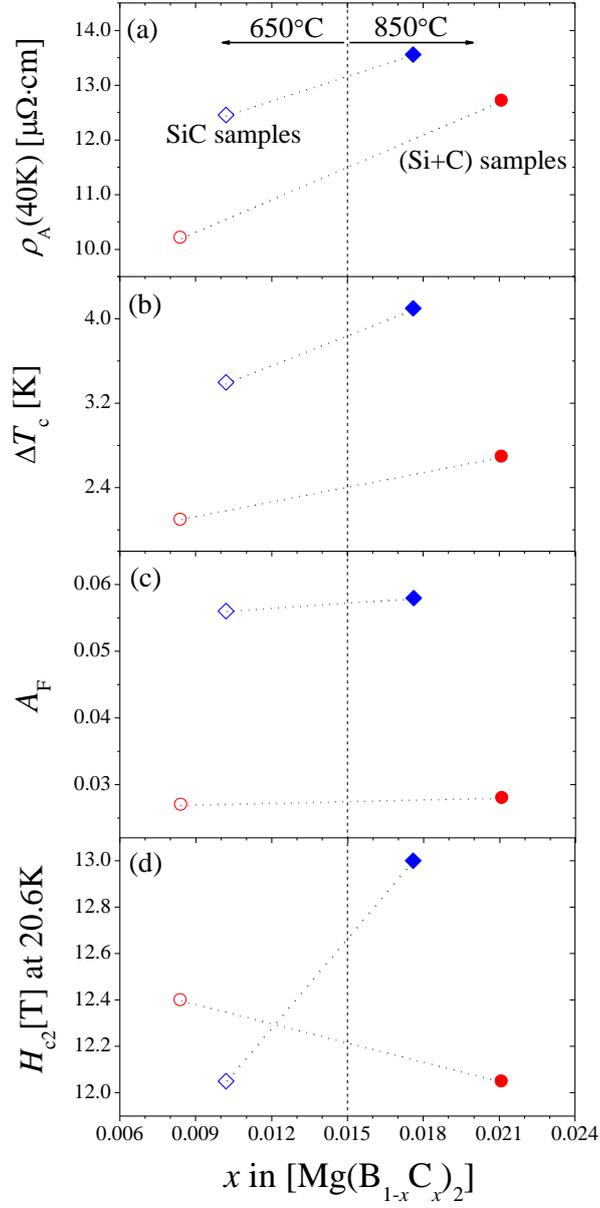

**Figure 5.** Variation of carbon substitution, $x$ with (a) corrected resistivity at 40K, $\rho_A(40K)$ (b) breadth of superconducting transition, $\Delta T_c$ (c) active area for supercurrent, $A_F$ and (d) upper critical field, $H_{c2}$. The Diamond and round symbols indicate SiC and (Si+C) samples, respectively. Dashed line separates the samples sintered at 650°C and 850°C. Dotted lines are given as guides to the eye only.

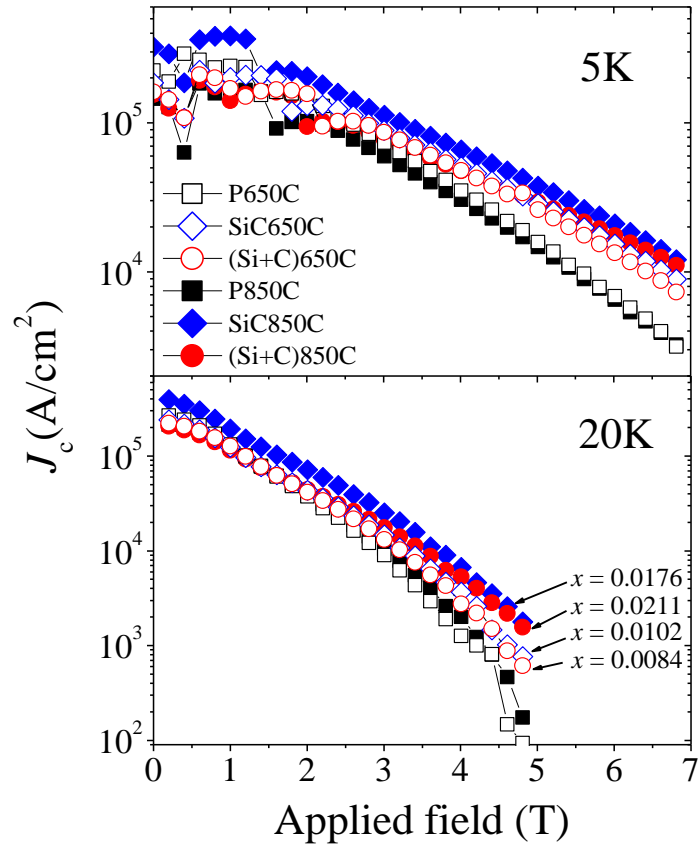

**Figure 6.** Dependency of critical current density, $J_c$, on applied magnetic field at 5 and 20K.